# Imaging pinning and expulsion of individual superconducting vortices in amorphous MoSi thin films

L. Ceccarelli, D. Vasyukov, M. Wyss, G. Romagnoli, N. Rossi, L. Moser, and M. Poggio

Department of Physics, University of Basel, 4056 Basel, Switzerland

Abstract

*We use a scanning nanometer-scale superconducting quantum interference device (SQUID) to image individual vortices in amorphous superconducting MoSi thin films. Spatially resolved measurements of the magnetic field generated by both vortices and Meissner screening satisfy the Pearl model for vortices in thin films and yield values for the Pearl length and bulk penetration depth at 4.2 K. Flux pinning is observed and quantified through measurements of vortex motion driven by both applied currents and thermal activation. The effects of pinning are also observed in metastable vortex configurations, which form as the applied magnetic field is reduced and magnetic flux is expelled from the film. Understanding and controlling vortex dynamics in amorphous thin films is crucial for optimizing devices such as superconducting nanowire single photon detectors (SNSPDs), the most efficient of which are made from MoSi, WSi, and MoGe.*

Introduction

The dynamics of quantized vortices play a crucial role in determining the electronic properties of devices made from type-II superconductors. The dissipative motion of vortices, driven by flowing electrical current, destroys the material's ability to carry current with zero resistance. By controlling the pinning of vortices, they can be immobilized, thereby restoring the superconducting state. Optimized pinning [1–9] and other methods of arresting vortex motion [10] have been used to extend the coherent dissipation-free superconducting state to high critical current densities, fields, and temperatures. The microwave response of superconductors is also affected by the high-frequency dynamics of vortices [11]. Vortex trapping contributes to loss in microwave resonators and conditions the performance of the circuits used as qubits in superconducting quantum computers [12]. Evidence has also emerged that superconducting vortices play a significant role in superconducting nanowire single photon detectors (SNSPDs) [13], whose high speed, detection efficiency, and low dark count rates make them attractive for a wide variety of applications. In particular, vortices are likely involved in both the mechanism used for the detection of photons and in the generation of dark counts.

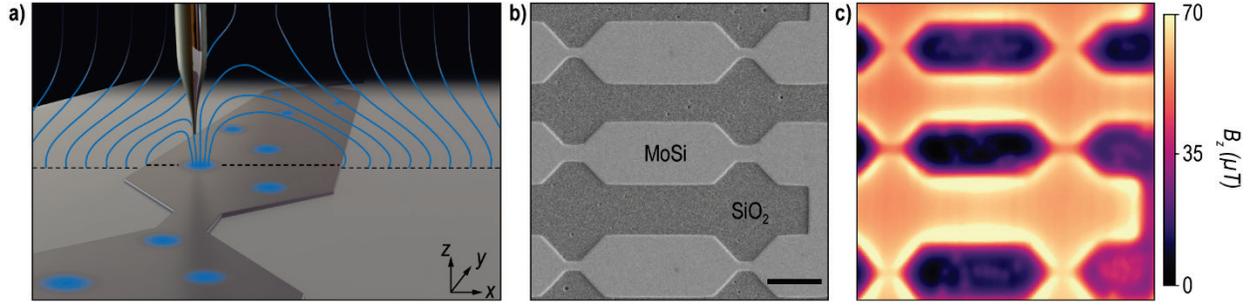

*Figure 1: Setup and MoSi sample. a) Schematic diagram of scanning SOT tip above superconducting MoSi wire with vortices. b) Scanning electron micrograph (SEM) and c) scanning SQUID image of the investigated region. Color-scale contrast in c) corresponds to the out-of-plane magnetic field B$_z$(x,y) measured by the SOT in an out-of-plane applied field of B$_a$ = 1.5 mT. Dark areas of low B$_z$ are due to the Meissner screening of the MoSi film, while bright dots indicate the penetration of flux in the form of superconducting vortices. Scale-bar: 10 μm.*

Here, we image individual superconducting vortices in thin-film MoSi using a scanning nanometer-scale superconducting quantum interference device (SQUID). Spatial maps of the magnetic field above patterned MoSi wires are collected as a function of applied field and reveal the presence of individual vortices and their pinning sites. Measured field profiles fit well to the Pearl model and yield the bulk London penetration depth $\lambda_L = 510 \pm 10$ nm at 4.2 K, which agrees with previous transport measurements of similar films [14]. Pinning sites of various strengths strongly affect the configuration of vortices. In particular, upon the reduction of the applied field, they hinder the expulsion of magnetic field from the film, resulting in long-lasting metastable vortex configurations. The results suggest that improved control of the density and strength of these pinning sites could be important for the optimization of devices based on amorphous MoSi films.

### Experimental setup

We investigate a $65 \pm 5$-nm-thick film of amorphous Mo$_{0.76}$Si$_{0.24}$ which is patterned into 9-μm-wide strips with short sections of narrower width, as shown in Fig. 1 b). Similar MoSi nanowires have recently been used as SPSPDs with some of the highest reported detection efficiencies [15]. The composition and thickness are chosen in order to maximize the critical temperature $T_c$, while obtaining an amorphous film with low defect densities and weak pinning effects [14], as desired in most applications.

We use a scanning SQUID-on-tip (SOT) sensor [16,17] to map the component of the magnetic field perpendicular to the MoSi film as a function of position and applied field at 4.2 K. The SOT used here has an effective diameter of 310 nm, as extracted from measurements of the critical current $I_{SOT}$ as a function of a uniform magnetic field applied perpendicular to the SQUID loop $\boldsymbol{B_a} = B_a \hat{\boldsymbol{z}}$ (see supplementary material). The MoSi sample is mounted below the SOT sensor in a plane parallel to the SQUID loop and perpendicular to $\boldsymbol{B_a}$. Since the SOT's current response is proportional to the magnetic flux threading through it, it provides a measure of the $z$-component of the local magnetic field $B_z$ integrated over the loop at its apex. By scanning the sample below the SOT at constant tip-sample spacing $z$, we map the local magnetic field $B_z(x, y)$. The sub-μm resolution is limited by the spacing and – ultimately – the SOT diameter.

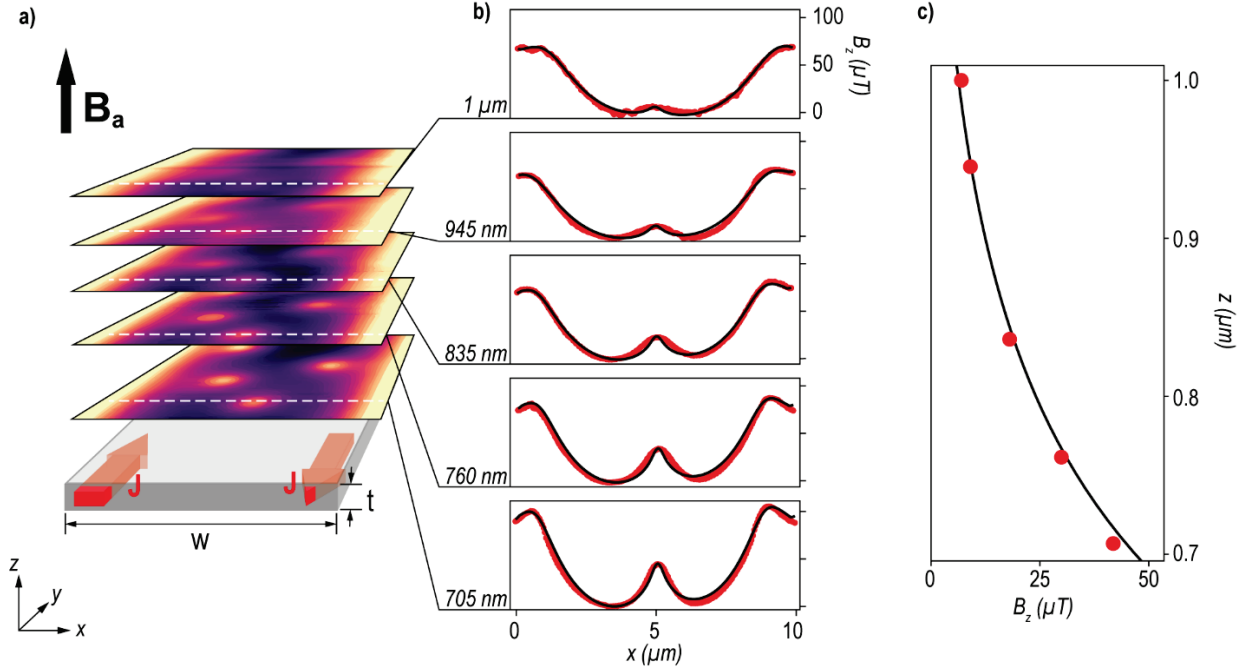

*Figure 2: Fits to Pearl and Meissner models. a) Schematic drawing of the MoSi wire with 2D measurements of $B_z(x,y)$ and b) line-cuts of $B_z(x)$ through a single vortex at different z (labelled) after field-cooling at $B_a$ = 0.5 mT. c) Magnetic field directly above the vortex as a function of the tip-sample distance z: $B_z(z)$. Red dots in b) and c) represent measurements and black lines fits (see supplementary material).*

### Vortex imaging

Before imaging $B_z(x,y)$ generated by the film, we nucleate the vortices by cooling the sample through its measured $T_c = 7.2$ K down to 4.2 K in an externally applied field $\boldsymbol{B_a}$. This field-cooling process causes flux penetrating the film above $T_c$ to concentrate in form of quantized vortices, which are trapped within the sample as it cools. The trapping potential is the result of the geometrical barrier produced by extended Meissner screening currents flowing near the sample edges [18]. Fig. 1(c) is a typical wide-field image of $B_z(x,y)$ at a fixed tip-sample spacing of $z = 705$ nm, showing the Meissner screening (dark regions) and the presence of a few trapped vortices (bright spots).

In Fig. 2(a), more detailed maps of $B_z(x,y)$ taken after field-cooling at $B_a = 0.5$ mT show a few isolated vortices. Line-cuts across the MoSi wire through the center of a single vortex are shown for different $z$ in Fig. 2(b) and the dependence of the field as a function of $z$ directly above the vortex is shown in Fig. 2(c). These data are well-fit by a model considering both the Meissner screening of the film [19] and the field produced by a perpendicular vortex in a film such that its thickness $t \ll \lambda_L$ (thin film limit), as first described by Pearl [20,21] (see supplementary material). Fits to these two effects measured at different $z$ allow us to independently determine the Pearl length $\Lambda = 2\lambda_L^2/t$ – and thus $\lambda_L$ – as well as the real tip-sample spacing $z$ for each measurement. At 4.2 K, we find $\Lambda = 8.1 \pm 0.9$ μm and a bulk $\lambda_L = 510 \pm 10$ nm, which agrees with measurements by Kubo of films with similar concentration and thickness [14].

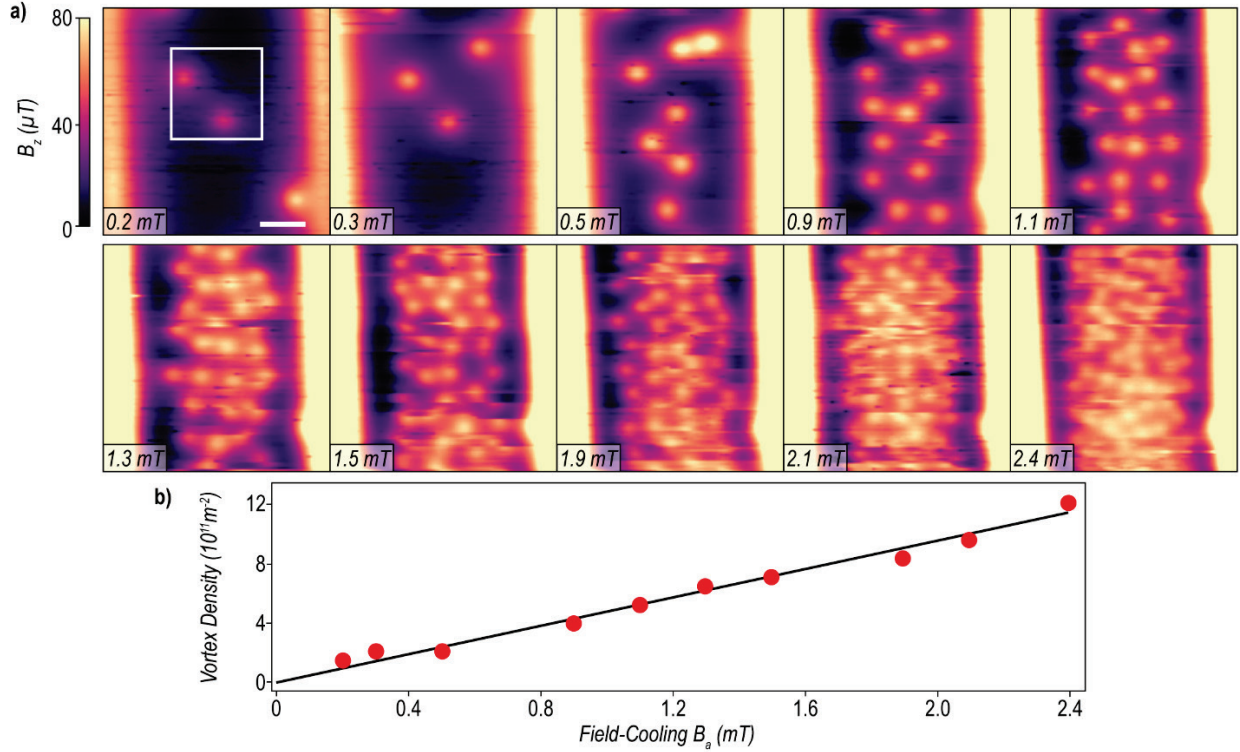

*Figure 3: Vortex density as a function of applied field. a) Maps of $B_z(x,y)$ taken at z=705 nm over a MoSi wire segment after field-cooling with increasing $B_a$, shown in the bottom left corner of each image. Scale-bar: 2 μm. b) Vortex density in a 4×4 μm area at the center of the MoSi wire, delineated by the white box in a), plotted as a function of the field-cooling field $B_a$. Data are shown as red dots, while the expected $B_a/\Phi_0$ dependence is shown as a black line.*

Vortex trapping and expulsion

By field-cooling with different values of $B_a$, we initialize different vortex densities in the MoSi film, as shown in Fig. 3. In the central region of the widest part of the wire, the measured vortex density depends linearly on $B_a$ with slope given by $\Phi_0^{-1}$, where $\Phi_0 = h/(2e)$ is the magnetic flux quantum, $h$ is Planck's constant, and $e$ is the elementary charge. These results agree with theoretical predictions [22] as well as previous measurements in similar Nb [23], YBCO [22], and Pb [24] wires. This behavior corresponds to the vortex density expected when the total flux through the film above $T_c$ nucleates into vortices and is trapped within the film. However, when the entire area of the MoSi wire is considered – including the edges – the measured vortex density is significantly less than the full above-$T_c$ flux density. The images of $B_z(x,y)$ in Fig. 3(a) show vortex-free regions near the sample edges, which shrink with increasing $B_a$, as also observed in Pb films by Jelić et al [25]. This non-uniform vortex density, which is concentrated in the center of the wire, reflects the reduced effective width, compared to the wire width, in which vortices can be trapped by the Meissner screening currents. In thin film wires whose width $w \gg t$, the screening current density decreases slowly from the edges, pushing vortices into the central part of the wire [18,26,27,25]. As a result, flux threading the sample near its edges above $T_c$ is subsequently expelled upon cool-down.

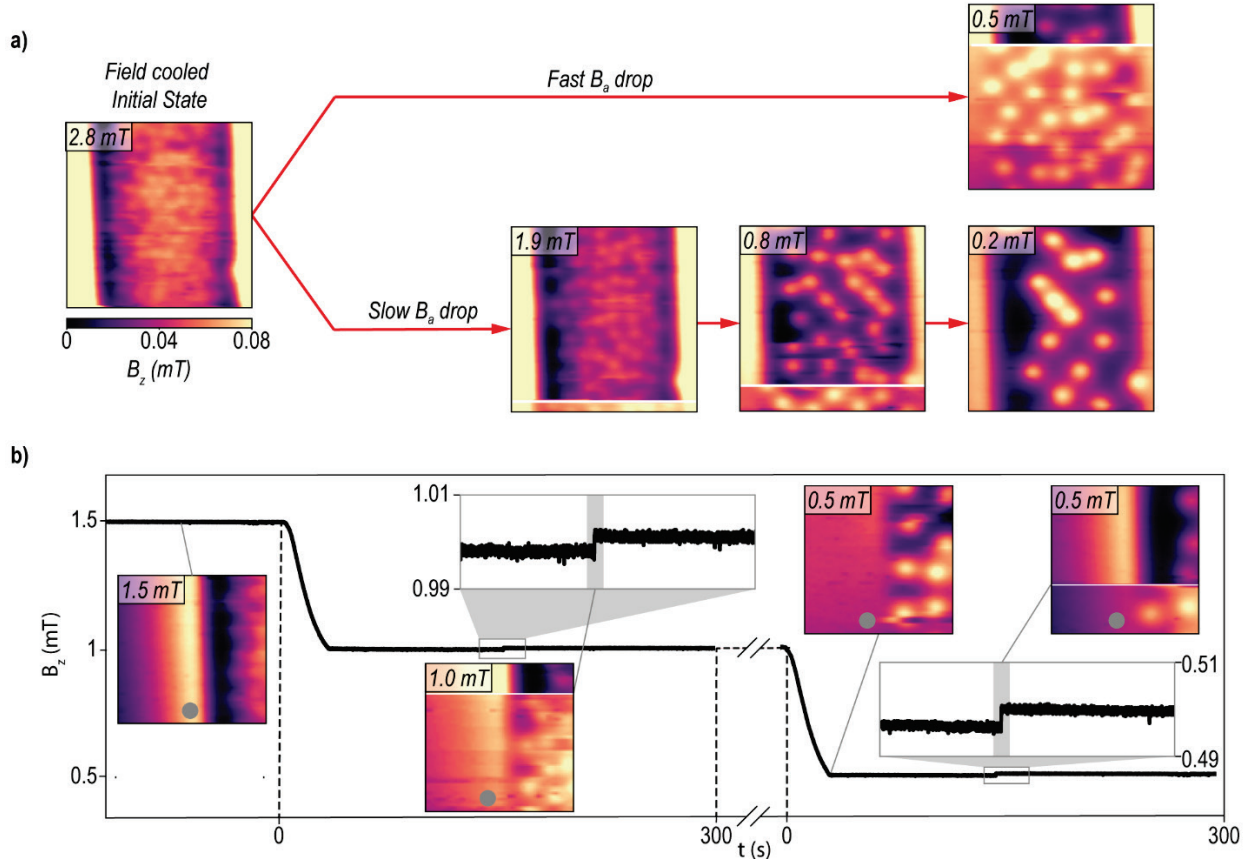

*Figure 4: 'Inflated' metastable vortex configurations. a) Left image shows a map of $B_z(x,y)$ 705 nm over a MoSi wire segment after field-cooling at $B_a$ = 2.8 mT. The image on the top right shows a measurement of the same area immediately after a decrease of $B_a$ to 0.5 mT in 30 s. The 10×10-μm² images are taken line-by-line in a total time of 260 s with x as the fast axis (71 lines at 3.7 s/line). The white line represents the moment at which we observe a discontinuous change in the vortex configuration. The bottom right images are taken at three fields as $B_a$ is reduced from 2.8 to 0.2 mT. b) 8.25×8.25-μm² images of $B_z(x,y)$ near the wire edge each taken in a total time of 245 s (51 lines at 4.8 s/line). The left-most image shows the configuration after field-cooling at $B_a$ = 1.5 mT. Other images show subsequent configurations after decreasing $B_a$ to 1 mT and then to 0.5 mT, as indicated in the top left corner of the images. The main plot shows $B_z$ as a function of time at the position indicated by the gray circle, showing both the reduction in $B_a$ and the signatures of vortex expulsion, which are highlighted in the zoomed-in sections. Note that the flux measured at this position (gray circle) and the spatial maps do not correspond to simultaneous measurements; they correspond to measurements carried out under the same initialization conditions. Therefore, the gray lines connecting the 2D maps to various points along the plot are meant to indicate similar events in different experimental runs, not two measurements of the same event.*

In order to investigate the potential trapping the vortices, after field-cooling, we image the vortex configuration upon the reduction of $B_a$. The left panel of Fig. 4 (a) shows an image of $B_z(x, y)$ after field-cooling at $B_a = 2.8$ mT to 4.2 K. The panels on the right show the same region immediately after ramping the applied field down to lower $B_a$ at constant temperature. In the case show in the top half of Fig. 4 (a), the field is ramped down $B_a = 0.5$ mT in 30 s and the $10 \times 10$-μm² image is taken line-by-line in a total time of 260 s with $x$ as the fast axis (71 lines at 3.7 s/line). Towards the end of the image, after 200 s, we observe a sudden change of the vortex configuration. The system goes from an 'inflated' state, in which vortices appear in the previously unoccupied edge-region, to a state similar to that observed after field-cooling, containing fewer trapped vortices and vortex-free edges. Upon further repeated imaging, the second configuration is always observed, except for small changes due to thermally activated vortex hopping, which will be discussed in the next section. In the bottom half of Fig. 4 (a), the field is reduced in three steps to $B_a = 0.2$ mT, with similar discontinuities marking a transition between 'inflated' and equilibrium states showing up in the images taken at $B_a = 1.9$ and 0.8 mT. The panels of Fig. 4 (b), showing a $8.25 \times 8.25$-µm² view of $B_z(x,y)$ near the wire edge (51 lines at 4.8 s/line), display similar behavior 190 s after $B_a$ is reduced from 1.5 to 1.0 mT and 320 s after it is subsequently reduced to 0.5 mT. In particular, we note that in the 'inflated' states observed at 1.0 and 0.5 mT, the vortices in the edge region appear pinned to the same set of pinning sites. Such vortex expulsion behavior is observed repeatedly for similar intervals of $B_a$ and over similar time-scales at constant temperature (see supplementary material).

To confirm that transitions from the 'inflated' state to the final state are not induced by interactions with our scanning tip [28], we carry out the same measurements with the SOT at a fixed position, just outside the edge of the sample, as marked by the gray circles in Fig. 4(b). Once again, $B_a$ is reduced at a constant temperature and the resulting field $B_z$ next to the wire is plotted as a function of time in the main part of Fig. 4(b). After the initial reduction due to the ramping down of $B_a$, $B_z$ is constant until around 150 s, when it increases in a sudden step. This increase, corresponds to the expulsion of the extra vortices present in the 'inflated' state. Vortex expulsion with similar time-scales is observed upon repeated experiments and a for different fields (see supplementary material). Since this behavior corresponds to what was observed in the scanning experiments, we rule out interactions with the tip as the trigger for the observed flux expulsion. Although strong interactions were previously observed between vortices in YBCO and a magnetic force microscopy tip [28], the much weaker stray fields produced by our SOT tip produce a negligible perturbation to each vortex, calculated to be less than 10 fN following the procedure described by Embon et al [24].

The observation of a metastable 'inflated' state, which lasts on the order of a hundred of seconds before flux expulsion and relaxation to the equilibrium configuration, points to the presence of pinning sites and thermally activated vortex hopping. The presence of vortices near the edges in the metastable state cannot be accounted for by the trapping potential provided by Meissner currents: this potential only traps vortices in the middle of the wire, with its barrier occupying the region near the edges. The vortices near the edges can only be immobilized by pinning sites. These potential wells, however, are not deep enough to completely freeze the vortex motion at 4.2 K and, due to thermally activated vortex hopping, the edge-vortices are eventually expelled as the sample relaxes to its equilibrium configuration.

Pinning sites

The presence of thermally activated vortex hopping at 4.2 K is evident from repeated imaging of $B_z(x,y)$ at equilibrium. As shown in Fig. 5 (a), certain vortices are seen to jump between neighboring pinning sites within a single image or between successive images. The former effect is evident in the discontinuities between one $x$-line and the next, while the latter can be seen when a vortex changes position between images. These jumps occur on a timescale ranging from a few to a few hundred seconds. Given this timescale and the similar timescale of the metastable configuration held together by pinning sites at the sample edges, we estimate the depth of these pinning potentials $E_{pin}$ using the Boltzmann formula for a thermally activated hopping rate: $\nu_{hop} = \nu_0 e^{-\frac{E_{pin}}{k_B T}}$, where $k_B$ is the Boltzmann constant, $T$ is the temperature, and $\nu_0$ is the attempt frequency. Solving for $E_{pin}$, we find an energy between 4 and 12 meV, whose uncertainty results from the imprecisely known ratio of $\frac{v_0}{v_{hop}}$. These weak pinning sites are therefore not strong enough to freeze vortex motion at 4.2 K and thus provide a path through which the system can eventually find its equilibrium configuration. As $B_a$ is increased, we note that pinning sites displaying thermally activated hopping occur more frequently from 25% of pinning sites at $B_a = 0.6$ mT to up to 85% at $B_a = 1.5$ mT (see supplementary material). As the vortex density increases, stronger vortex-vortex interactions, which have a long range $1/r$ force in thin films, destabilize an increasing number of vortex pinning sites.

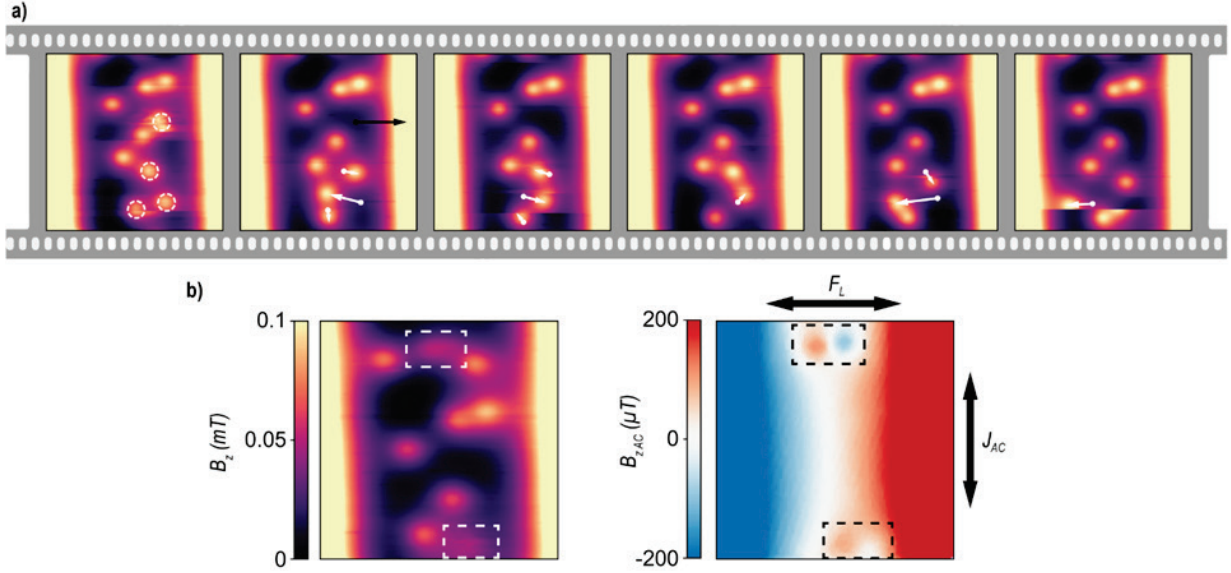

*Figure 5: Thermally activated vortex hopping and pinning strength. a) Maps of $B_z(x,y)$ at z=705 nm over a MoSi wire segment after field-cooling with $B_a$ = 0.8 mT. Each 10×10-µm² image is of the same region and is taken one after the other line-by-line in 260 s with x as the fast axis (71 lines at 3.7 s/line). Circles and arrows highlight vortices hopping between neighboring pinning sites. b) Left image is a Map of $B_z(x,y)$ under the same conditions, but in the additional presence of an AC applied current along $\hat{y}$, producing a FAC in $\hat{x}$. The right image shows $B_{z,AC}(x,y)$, which is measured simultaneously. The highlighted areas show vortices which oscillate as a result of $F_{AC}$. Scan area: 10x10-µm².*

Although many pinning sites exhibit thermally activated vortex hopping, some sites do not, pointing to deeper trapping potentials. We further investigate these sites, by driving vortex motion using an AC current $I_{AC}$ applied along the $y$-direction (long direction of the wire). This current results in an AC Lorentz force $F_{AC} = \frac{I_{AC}}{w}\Phi_0$ on each vortex along $x$-direction. $F_{AC}$ drives oscillations of the vortices around their equilibrium positions. The resulting motion can be measured using the scanning SOT as either a blurred vortex image in $B_z(x,y)$, as shown in the left panel of Fig. 5 (b), or – more clearly – as a dipolar signal in the magnetic field produced by the sample at the frequency of the AC current drive $B_{z,AC}(x,y)$, as shown in the right panel [24]. The amplitude of this signal reflects the size of the oscillation and the gradient of the static field $B_z(x,y)$ produced by the vortex.

Vortices at different pinning sites respond differently to $F_{AC}$, indicating the varying strength of the pinning potentials throughout the sample. At $B_a = 0.9$ mT, weak pinning sites (30%), showing thermally activated hopping, are induced to jump from site to site by $F_{AC}$ with amplitudes in the 100 fN range. Strong pinning sites (50%) show no measureable response to this $F_{AC}$. Intermediate pinning potentials show oscillatory responses to the $F_{AC}$. From the strength of $F_{AC}$, we extract effective pinning spring constants of around 0.6 µN/m. Compared to similar measurements in Pb films, these potentials are 100 times shallower (less curvature) indicating their larger spatial extent and weaker strength. By increasing Lorentz force up to a maximum value $F_{AC,max}$ at which these vortices escape from their pinning sites, we can integrate and estimate the depth of the potential well, typically obtaining $E_{pin} \sim 60$ meV.

Together, our results point to a range of $E_{pin}$ from a few meV up to at least tens of meV, excluding the pinning sites that do not respond to thermal or Lorentz perturbations. As expected, these values are a fraction of the estimated condensation energy of the vortex core in this sample of $\frac{H_c^2}{2\mu_0}\pi\xi^2 t = 110$ meV, where $\xi = 7$ nm is the coherence length derived from a measurement of the upper critical field $H_{c2} = \frac{\Phi_0}{2\pi\xi^2} = 6.8$ T [14], $H_c = \frac{\Phi_0}{\sqrt{8}\pi\lambda_L\xi} = 66$ mT is the thermodynamic critical field, and $\mu_0$ is the permeability of free space.

If we exclude the weak thermally activated pinning sites, we also find that the vortices are always driven in phase with $F_{AC}$ and along the direction of the drive ($x$-direction). This behavior points to isotropic conservative trapping potentials, which are sparsely distributed, i.e. large pinning centers that do not overlap. These large and well-separated pinning centers contrast the observations made by Embon et al. in similar experiments on Pb thin films [24]. There, vortices displayed anisotropic responses and spring softening upon driving with Lorentz forces. Those data were well-described by a computational model that included small clusters of pinning defects separated on the order of a superconducting coherence length.

The vortex pinning potentials observed in our MoSi films are likely the result of either crystalline precipitates within the amorphous alloy or structural and surface defects. The latter cause appears the most probable, given the presence of visible inclusions and roughness of varying size in SEMs of our sample (see supplementary material). Precipitates are typically a limiting factor in much thicker films, which require long sputtering times [14].

Conclusions

Our scanning SOT experiments, because of the sensors combination of high spatial resolution and high magnetic field sensitivity, reveal images of individual superconducting Pearl vortices in amorphous MoSi thin films. In addition to providing a measure of the penetration depth of the film, we directly observe their thermally activated hopping at 4.2 K. Since the vortices are not completely frozen onto their pinning sites, we are able to estimate the depth of the weakest pinning potentials and observe metastable vortex configurations present for tens of seconds before the system reaches equilibrium.

The dynamic nature of the vortex configurations in MoSi at 4.2 K may have implications for SNSPDs and other devices fabricated from such films. The thermally activated vortex hopping observed here may be a source of residual dark counts in SNSPDs. Our experiments also make plain the necessity of further reducing the density of unintentional pinning sites by improving sample quality. In order to aid in the optimization of MoSi-based superconducting devices, future scanning SQUID studies should aim to study different MoSi films grown under different conditions.

Methods

**MoSi sample fabrication.** The $65 \pm 5$-nm-thick $Mo_{0.76}Si_{0.24}$ film is deposited onto an $SiO_2$ substrate by co-sputtering with a DC and RF bias on Mo and Si targets, respectively. The thickness is determined using atomic force microscopy (AFM) and SEM, while the concentration of Mo and Si are measured by energy-dispersive x-ray (EDX) spectroscopy (see supplementary material). The film is patterned into a series of meandering wires over a $500 \times 500$-μm$^2$ area (see supplementary material) by a combination of electron-beam lithography and reactive ion etching.

**SOT sensor fabrication.** The SOT is fabricated by evaporating Pb on the apex of a pulled quartz capillary according to a self-aligned method pioneered by Finkler et al. [16] and perfected by Vasyukov et al [17]. The evaporation is carried out in a custom-made evaporator with a base pressure of 2 × 10$^{-8}$ mbar and a rotatable sample holder cooled by liquid He. In accordance with Halbertal et al. [29], an additional Au shunt was deposited close to the tip apex prior to the Pb evaporation for protection of the SOTs against electrostatic discharge. SOTs were characterized in a test setup prior to their use in the scanning probe microscope (see supplementary material).

**Scanning SOT experiments.** The SOT and the MoSi sample are mounted in a custom-built scanning probe microscope operating under vacuum in a $^4$He cryostat [30]. A serial SQUID array amplifier is used to measure the current flowing through the SOT (see supplementary material). Positioning and scanning of the sample below the SOT is carried out using piezo-electric positioners and scanners (Attocube AG). In order to aid in positioning the sensor over the area of interest at low temperature, we pass AC electric currents through the meandering wires patterned in the MoSi (see supplementary material). The SOT's sensitivity to the resulting Biot-Savart fields allows us to navigate across the sample. In addition, close to the surface, the distortion the local DC magnetic field by the Meissner effect helps to determine the sensor's position, e.g. as shown in Fig. 1 c). All measurements were performed in open-loop mode, at constant spacing above the sample.

Acknowledgements

We thank Prof. Christian Schönenberger, Prof. Richard Warburton, and Daniel Sacker for guidance in the fabrication of the MoSi samples. We also thank the machine shop of the Department of Physics and Dr. Monica Schönenberger and the Nano Imaging Lab for support with part design and AFM, respectively. We acknowledge the support of the Canton Aargau; the Swiss Nanoscience Institute; the ERC Starting Grant NWScan (Grant 334767); and the Swiss National Science Foundation via grant 200020-159893, grant 200020-178863, the Sinergia network Nanoskyrmionics (grant CRSII5-171003), and the NCCR Quantum Science and Technology (QSIT).

# Supplementary Material: Imaging pinning and expulsion of individual superconducting vortices in amorphous MoSi thin films

L. Ceccarelli, D. Vasyukov, M. Wyss, G. Romagnoli, N. Rossi, L. Moser, and M. Poggio

Department of Physics, University of Basel, 4056 Basel, Switzerland

## SQUID-on-tip characteristics

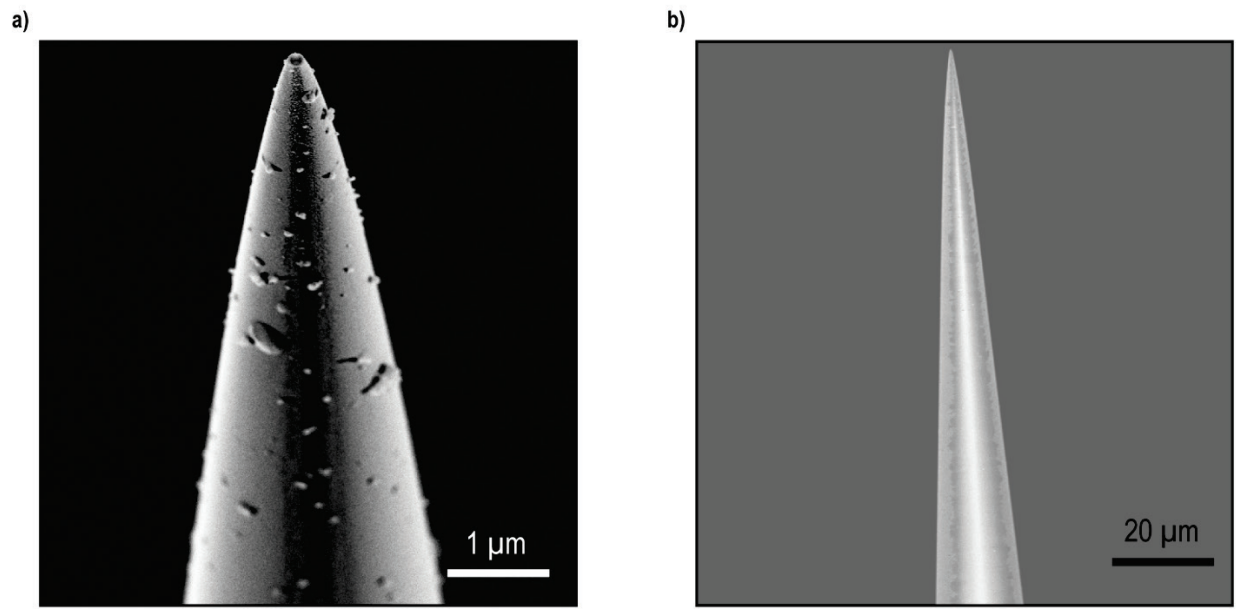

*Figure S1: SEMs of a typical SOT sensor. a) zoomed-in and b) zoomed-out views.*

The SQUID-on-tip (SOT) sensor used here is made by evaporating Pb on the apex of a pulled quartz capillary according to a self-aligned method described in Vasyukov et al [1]. A typical SOT is shown in the scanning electron micrographs (SEMs) of Fig. S1. The two evaporated Pb leads are visibly separated by an insulating gap. The zoomed-in SEM shows the size of the SQUID loop at the tip.

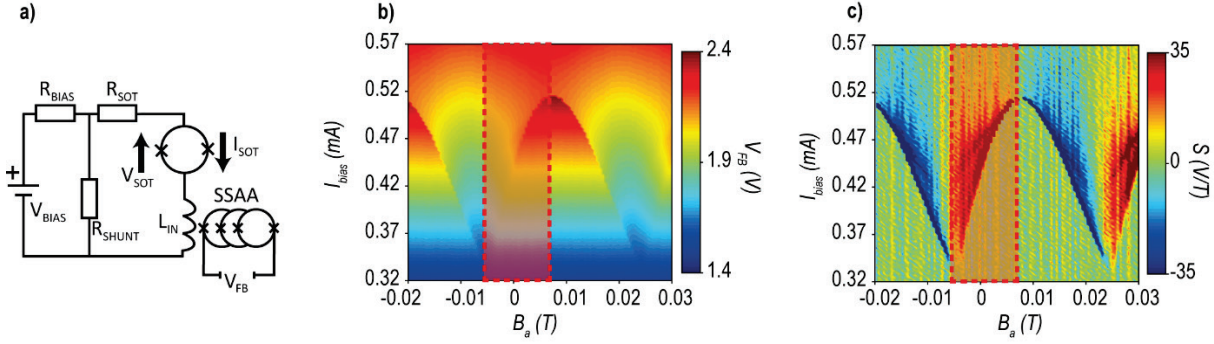

*Figure S2: SOT circuit and quantum interference. a) SOT measurement circuit diagram. b) Quantum interference of the SOT and b) its derivative with respect to $B_a$.*

The effective diameter of this SQUID loop can be determined from the quantum interfence pattern measured at $T$ = 4.2 K using the circuit shown in shown in Fig. S2 (a). The SOT shows pronounced critical current $I_c(B_a)$ oscillations, shown inf Fig. S2 (b) with a period corresponding to an effective loop diameter of 310 nm. A serial SQUID array amplifier (SSAA) is used to measure the current passing through the SOT $I_{SOT}$ yeilding an output voltage $V_{FB}$ proportional to $I_{SOT}$. The sensitivity of the SOT to magnetic field for various $I_{bias}$ and $B_a$ is shown in Fig. S2 (c) by plotting $\frac{\partial V_{FB}}{\partial B_a}$. This map is used to determine its sensitivity to magnetic field. Note that for certain values of $B_a$, the SOT is nearly insensitive to variations in magnetic field.

It is interesting to observe that the SQUID interference pattern of this particular device is not symmetric respect to $B_a$ = 0 T. This shift of the pattern is due to the asymmetry of the two Dayem bridge junctions. In our case, this asymmetry was produced intentionally by modifying the fabrication recipe, so that we could investiate the vortex state of the MoSi sample even for very small $B_a$.

Sample thickness and surface

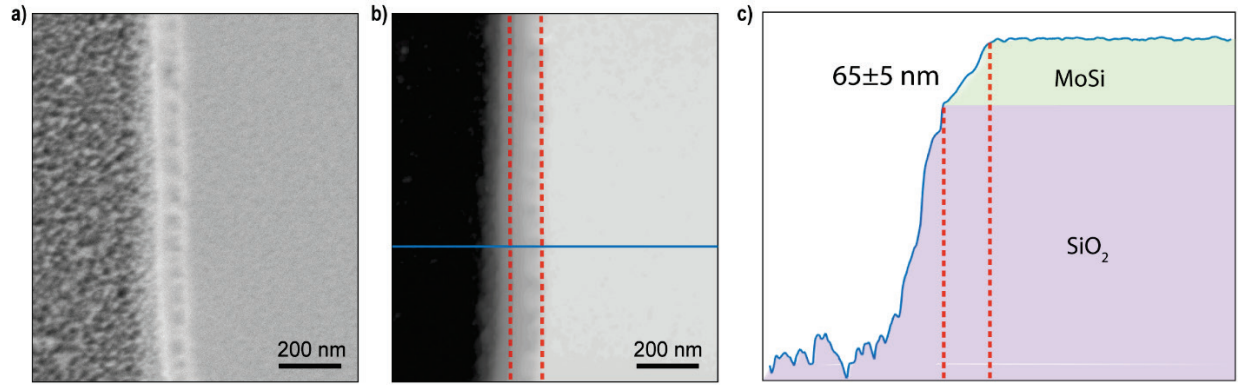

*Figure S3: SEM and AFM of MoSi sample. a) SEM of an edge of the MoSi wire. The dark region is SiO$_2$ and the light region is the MoSi. b) AFM image of the same region with dark contrast corresponding to low parts of the sample and light contrast to high parts of the sample. c) Line cut of the AFM in b).*

The thickness of the MoSi film is an important parameter, which determines whether our system resides in the superconducting thin-film limit and what kind model should be used to describe the penetrating magnetic flux. The thickness was measured both by atomic force microscopy (AFM) and SEM. Figs. S3 a) and b) show SEM and AFM measurements of the same area, showing the edge of a MoSi wire and the transition from etched SiO$_2$ to the MoSi thin-film. An AFM line-cut, taken along the blue line in Fig. S3 b) is depicted in Fig. S3 c). The two dashed red lines in Fig. S3 b) and c) denote the edge of the MoSi layer, whose thickness is highlighted in greem in Fig. S3 c).

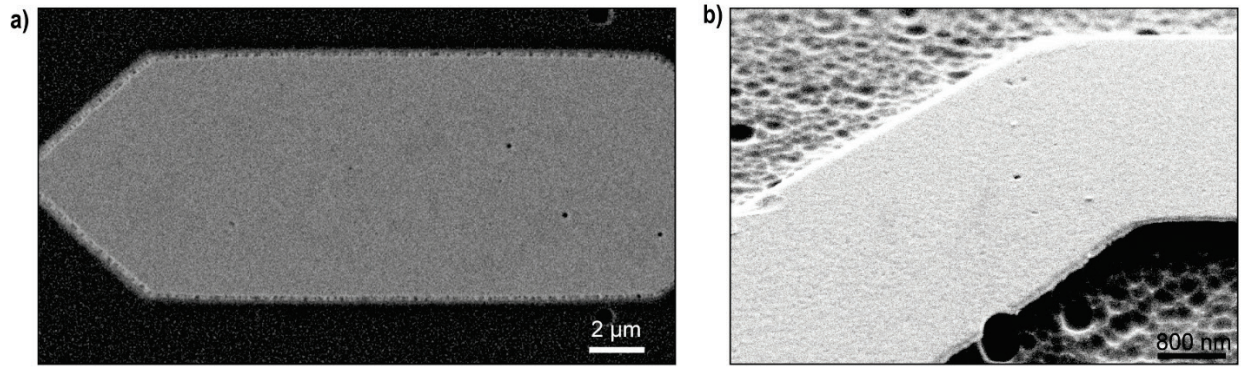

*Figure S4: SEMs of MoSi sample from the a) top and from b) an oblique angle.*

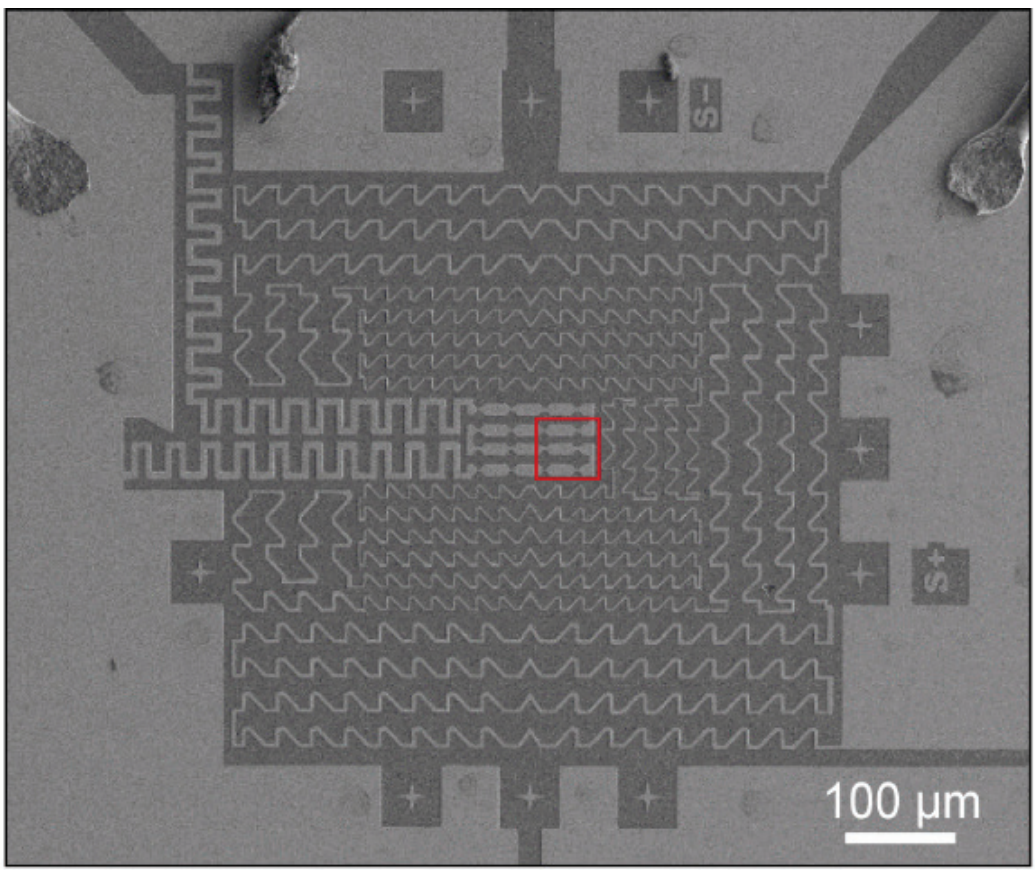

*Figure S5: Wide-field SEM of the MoSi sample.*

Structural and surface defects in the MoSi film are shown in Figs. S4 a) and b) as inclusions and roughness on the surface. These defects are especially prominent near the edges of the sample.

Wide-field SEM of the MoSi sample is shown in Fig. S5 a). The red-square highlights the area corresponding to the SEM and magnetic map reported in Figs. 1 b) and c), in the main text.

Model for fields generated by Meissner effect and Pearl vortices

The model used to fit the spatial dependence of $B_z$ measured above the superconducting MoSi wire takes into account the wire's expulsion of flux due to the Meissner effect and the field produced by a superconducting vortex in the Pearl limit. By summing the resulting field profiles, we model the 1D profile shown in Figs. 3 b) and c), which cut through the center of the field produced by a single vortex in the wire.

The model for the field produced by Meissner screening follows Brisbois et al. [2]:

$$B_{\text{Meissner}}(x,z) = 1 + \frac{1}{\pi} \int_{-\frac{w}{2}}^{\frac{w}{2}} \frac{(x-x')x'dx'}{[(x-x')^2 + z^2]\sqrt{\left(\left(\frac{w}{2}\right)^2 - x'^2\right) + \frac{8a\lambda_L^2}{\pi t}}}$$

The model for the field produced by the superconducting vortex is based on Pearl's model, because the wire thickness is much less than the penetration depth ($t \ll \lambda_L$) [3,4]:

$$B_{z,\text{vortex}}(x,z) = \Phi_0 \int_{-\frac{1}{L}}^{\frac{1}{L}} \frac{d^2k}{4\pi^2} \frac{e^{i\mathbf{k}\cdot\mathbf{x}}}{1+\frac{4k^2\lambda_L^4}{t^2}} f(k,z)$$

where $f(k,z) = c_1 e^{-kz}$ above the film, i.e. for $z > 0$, where 0 is the surface of the film. $c_1$ depends on the film thickness $t$:

$$c_1(k) = [(k+\rho)e^{\rho t} + (k-\rho)e^{-\rho t} - 2k]\frac{\rho}{c_2}$$

$$c_2(k) = (k+\rho)^2 e^{\rho t} - (k-\rho)^2 e^{-\rho t}$$

$$\rho = \sqrt{k^2 + \left(\frac{2\lambda_L^2}{t}\right)^{-2}}$$

The measured field is then fit with the function $B_{z,fit}(x,z) = B_{z,Meissner}(x,z) + B_{z,vortex}(x,z)$, where the London penetration depth $\lambda_L$ and probe-sample distance z are used as fit parameters. The value of $\lambda_L$ is found to be $510 \pm 10$ nm (making the Pearl penetration depth $\Lambda \cong 8$ μm) and the z values are shown in Fig. 3 b) for five different probe-sample distances.

Thermally activated vortex hopping as a function of vortex density

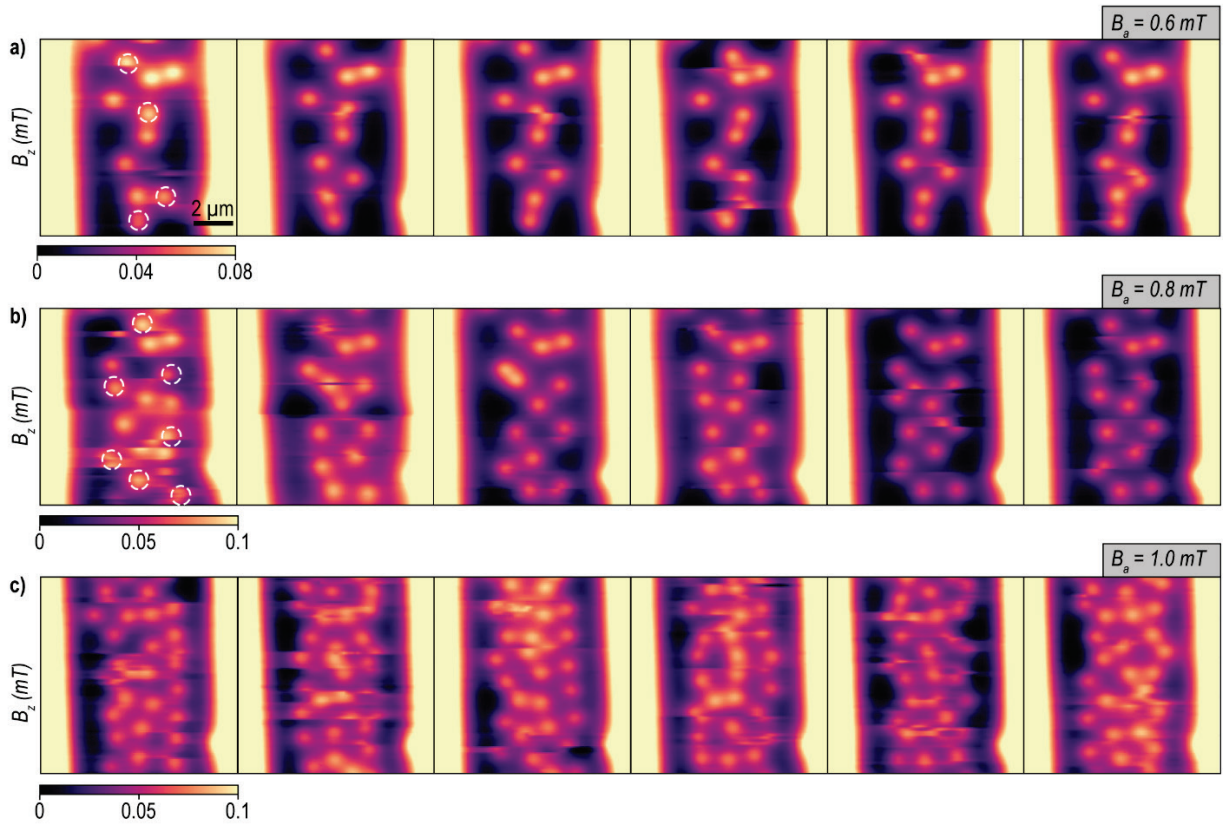

*Figure S6: Dependence of thermally activated vortex hopping on applied field. a), b), and c) show a time series of images of $B_z(x,y)$ taken at three different field-cooling fields $B_a$.*

In Fig. S6, we show 3 series of images taken one after the other for 3 different vortex configurations, initialized with Ba = 0.6, 0.8, and 1.0 mT, respectively. Each frame requires 153 s to measure and successive images are taken immediately after the previous image finishes. For the first two sequences a) and b), we highlight the initial positions of the moving vortices. For the third sequence c), the high vortex density makes almost all the vortices mobile.

Fig. S6 makes clear that, as $B_a$ is increased, the proportion of pinning sites showing thermally activated vortex hopping increases. This is likely due to the increasing vortex density and the resulting increase in the strength of vortex-vortex interactions.

For a clearer visualization, a series of supplementary movies of the vortex motion shown in Fig. S6 as well as additional measurements are available for download. The movies show successive images of under the same conditions as Fig. S6 for field-cooling fields $B_a$ = 0.6, 0.7, 0.8, 0.9, 1.0, and 1.5 mT with the names 0.6mT_Ba.avi, 0.7mT_Ba.avi, 0.8mT_Ba.avi, 0.9mT_Ba.avi, 1.0mT_Ba.avi, and 1.5mT_Ba.avi, respectively.

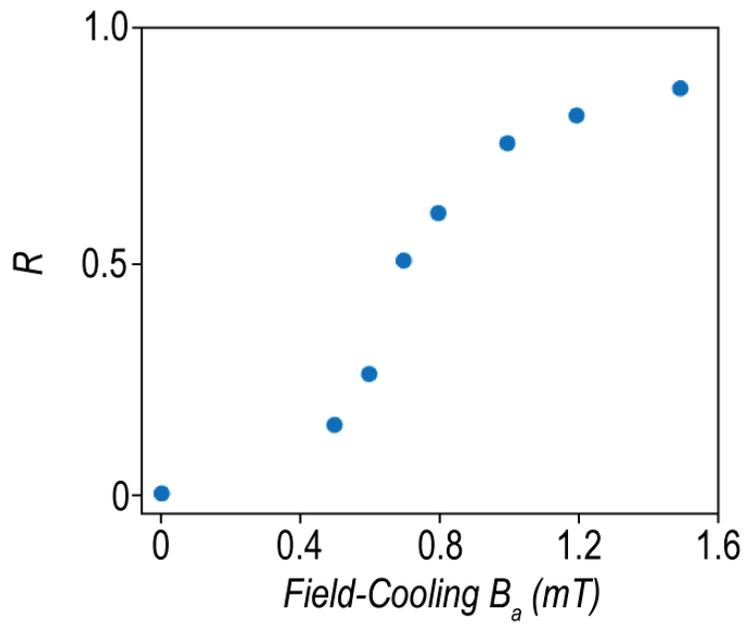

*Figure S7: Proportion vortices showing thermally activated hopping as a function of $B_a$.*

Figure S7 shows the relation between the proportion of vortices showing thermally activated hopping, $R = \frac{n_m}{n_t}$, and the field-cooling field $B_a$, where $n_m$ is the density of vortices showing thermally activated hopping and $n_t$ is the total vortex density. $R$ is seen to increase until nearly all vortices show thermally activated behavior.

EDX data on Si concentration

In order to confirm the high purity of the Mo-Si wafers, we perform X-ray photoelectron spectroscopy (XPS) characterization of MoSi film used to produce the wire samples (figure S8). Besides the two elements constituting the samples (Mo and Si), only oxygen and carbon are measured. Their presence is due to atmosphere contamination as the samples are exposed to air during the process. Two fine scans of the Mo3d and Si2p regions are done to identify the different chemical states and are fitted to provide the most accurate composition calculations as possible. Both Mo and Si are found in a MoSi alloy state (227.6 and 99 eV, respectively) and oxide state (232.2 and 102 eV, respectively), where the latter is expected due to air exposure [5]. In terms of elemental composition, the surface is composed of 24 at. % of Si for 76 at. % of Mo.

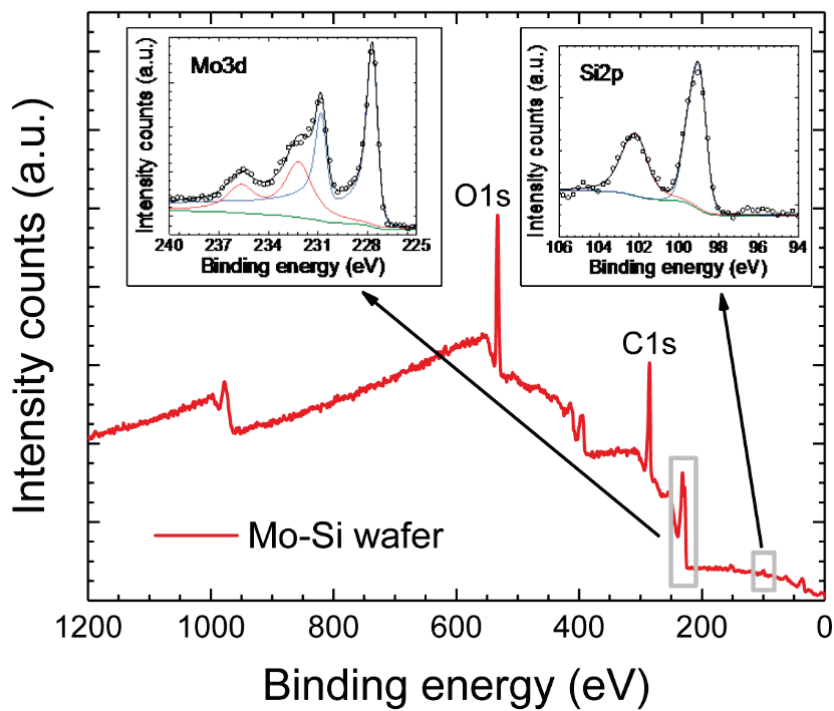

*Figure S8: XPS characterization of the MoSi film shown in red. Two narrow scans of the Mo3d and Si2p regions are shown as insets. The shown spectra are normalized for comparison. The open circles are the measured data and the black lines correspond to the sum curve of all components represented in colored lines.*

Images of vortex expulsion process

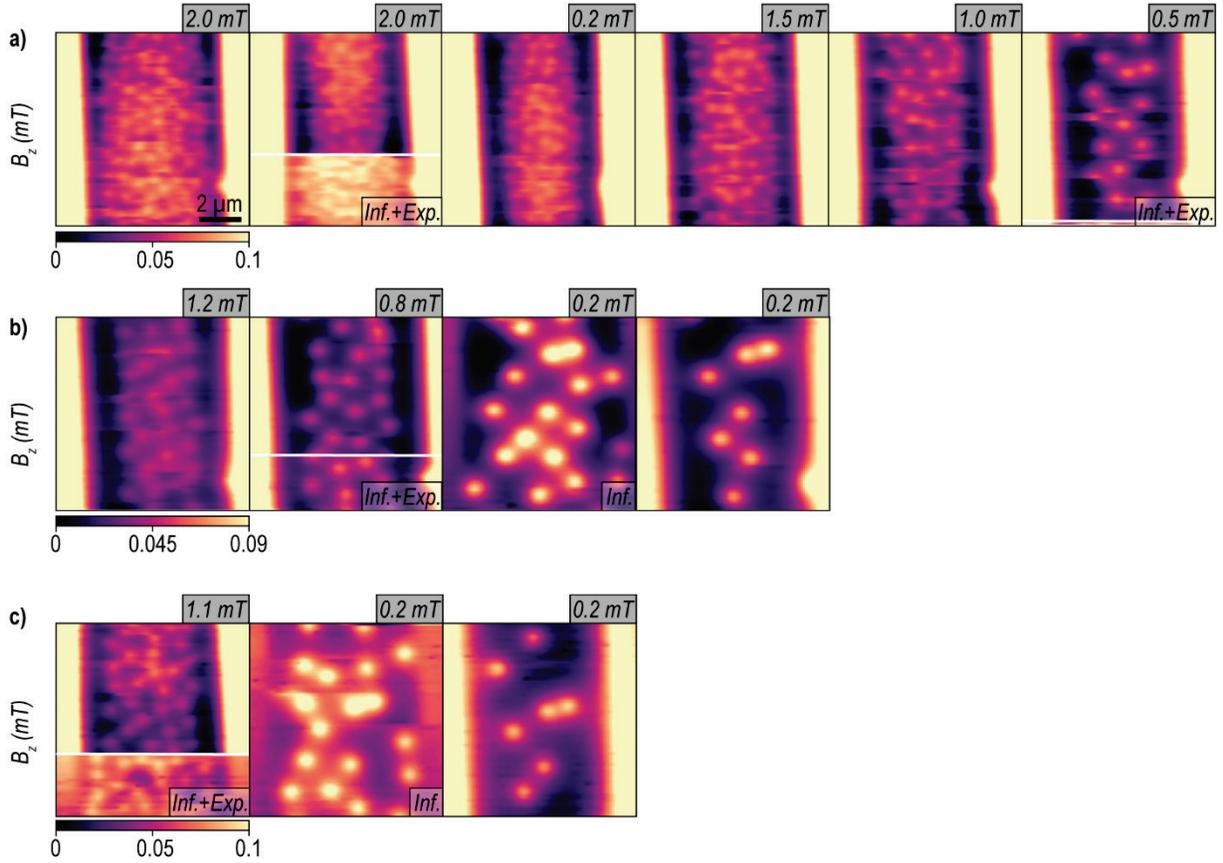

*Figure S9: Futher 'inflated' metastable vortex configurations. a), b), and c) show three series of successive images of $B_z(x,y)$ taken at the fields labeled in the upper left corners. The labels 'Inf.' and 'Exp.' indicate whether a scan shows 'inflated' states and whether an expulsion (indicated by the white horizontal line) occurs during the scan, respectively.*

In Fig. S9, we report other examples of metastable 'inflated' states followed by vortex expulsions, similar to those presented in Fig. 4 in the main text.